# Revisiting Relative Indicators and Provisional Truths

Loet Leydesdorff *[a] and Tobias Opthof [b]

**Abstract**

Following discussions in 2010 and 2011, scientometric evaluators have increasingly abandoned relative indicators in favor of comparing observed with expected citation ratios. The latter method provides parameters with error values allowing for the statistical testing of differences in citation scores. A further step would be to proceed to non-parametric statistics (e.g., the top-10%) given the extreme skewness (non-normality) of the citation distributions. In response to a plea for returning to relative indicators in the previous issue of this newsletter, we argue in favor of further progress in the development of citation impact indicators.

---

[a] *corresponding author; Amsterdam School of Communication Research (ASCoR), University of Amsterdam**,** PO Box 15793, 1001 NG Amsterdam, The Netherlands; loet@leydesdorff.net
[b] Experimental Cardiology Group, Heart Failure Research Center, Academic Medical Center AMC, Meibergdreef 9, 1105 AZ Amsterdam, The Netherlands; tobias.opthof@gmail.com

**Introduction**

In the *ISSI Newsletter 14*(2), Glänzel & Schubert (2018) argue for using "relative indicators" — e.g., the Mean Observed Citation Rate relative to the Mean Expected Citation Rate *MOCR/MECR* (Schubert & Braun, 1986; cf. Vinkler, 1986) —instead of testing citation scores against their expected values using the mean normalized citation score *MNCS* (Waltman, Van Eck, Van Leeuwen, Visser, & Van Raan, 2011a and b). The authors note our "concern" about using these relative indicators (Opthof & Leydesdorff, 2010; cf. Lundberg, 2007). However, Glänzel & Schubert (2018) state (at p. 47) that they do not wish to "resume the debate but attempt to shed some light on the premises and the context of indicator design in the mirror of the rules of mathematical statistics."

In their discussion of the indicators, Glänzel & Schubert (2018) pay insufficient attention to the differences in terms of the results of a scientometric evaluation. Are the indicators valid and reliable (Lehman *et al*., 2006)? Our "concern" was never about the relative indicators as mathematical statistics, but about their use in evaluations. From this latter perspective, the division between two averages instead of first normalizing against expected values can be considered as a transgression of the order of mathematical operations by which division precedes addition.

In the case of *MOCR/MECR*, one first sums in both the numerator and denominator and then divides, as follows:

$$\frac{MOCR}{MECR} = \frac{\sum_{i=1}^{n} c_i/n}{\sum_{i=1}^{n} e_i/n} = \frac{\sum_{i=1}^{n} c_i}{\sum_{i=1}^{n} e_i} \qquad (1)$$

In the case of *MNCS*, one first divides and sums thereafter:

$$MNCS = \frac{1}{n}\sum_{i=1}^{n}\frac{c_i}{e_i} \qquad (2)$$

Eq. 1 has also been called the "Rate of Averages" (RoA) versus the "Average of Rates" (AoR) in the case of Eq. 2 (Gingras & Larivière, 2011).

**The old "crown indicator"**

The "relative indicators" of Eq. 1 were introduced by the Budapest team in the mid-1980s (Schubert & Braun, 1986; Vinkler, 1986). One of these relative indicators—using the field of science as the reference set—has been used increasingly since approximately 1995 as the so-called "crown indicator" (*CPP/FCSm*)[3] by the Leiden unit *CWTS* (Moed, De Bruin, & Van Leeuwen, 1995). These "relative indicators" are still in use for research evaluations by the ECOOM unit in Louvain headed by Glänzel.

In a vivid debate, Van Raan *et al*. (2010) first argued that the distinction between RoA and AoR was small and therefore statistically irrelevant. However, both Opthof & Leydesdorff (2010) and

---

[3] *CPP/FCSm* is the total "citations per publication" for a unit under evaluation divided by the mean of the citations in the respective field.

Gingras & Larivière (2011) provided examples showing significant differences between the two procedures. Using AoR, one is able to test for the statistical significance of differences in citations among sets of documents. Unlike AoR, RoA comes as a pure number (without error); using this indicator at the time, CWTS and ECOOM invented "rules of thumb" to indicate significance in the deviation from the world standard as 0.5 (Van Raan, 2005) or 0.2 (CWTS, 2008, at p. 7; cf. Schubert & Glänzel, 1983; Glänzel, 1992 and 2010). Even if one tries to circumvent the violation of basic mathematical rules by adding brackets to the equations, these conceptual issues remain.

**AoR and RoA in the banking world**

Glänzel & Schubert (2018) refer to a paper published in the arXiv by Matteo Formenti (2014) from the Group Risk Management of the UniCredit Group. In this risk assessment, the author compares default rates of mortgages issued in the years 2008-2011 during the subsequent five years as risks for the bank. The time of default applies to any mortgage that ends before the scheduled date planned by the bank, either because the individual fails to pay or because the mortgage is paid off before the planned date, which also implies less income for a portfolio holder such as a bank.

The problem formulation is different from that of research evaluation using citations:

1. For a bank it does not matter which customers fail to pay the mortgage in the future, as long as the sumtotal of individual positions of customers does not provide a risk for the bank. The sumtotal provides the reference in RoA;
2. Formenti (2014) missed an important issue: in his test portfolio there are 12 risk groups from 'M1' to 'M12', with the highest risk residing in 'M12'. Neither RoA nor AoR are able to estimate the risk in the highest risk group or the risk groups with a lower but still substantial risk profile; both indicators underestimate the risk by an order of magnitude. Analogously, the risks in the lowest risk group ('M1') are grossly overestimated, regardless of whether RoA or AoR is used. (Because both estimates thus fail, holders of home mortgages pay an interest rate on loans much higher than the current one on the market.)

We do not understand the relation between this example and research evaluations. Are funding agencies distributing money over the scientific community with the aim of avoiding their own bankrupcy?

**The new "crown indicator"**

In the weeks after voicing our critique (in 2010), the Leiden unit turned up another "crown indicator:" *MNCS* or the "mean normalized citation score" (Eq. 2; Waltman, van Eck, van Leeuwen, Visser, & van Raan, 2011 a and b). In our response, we expressed our concern about moving too fast—without sufficient debate—to this alternative (Leydesdorff & Opthof, 2011). Following up on Bornmann & Mutz (2011), we then proposed "to turn the tables one more time" by first specifying criteria for comparing sets of documents in terms of performance indicators independently from specific evaluation contexts and existing infrastructures (Leydesdorff, Bornmann, Mutz, & Opthof, 2011). We formulated these criteria (at pp. 1371f.), as follows:

1. A citation-based indicator must be defined so that the choice of the reference set(s) (e.g, journals, fields) can be varied by the analyst independently of the question of the evaluation scheme. In other words, these two dimensions of the problem (the normative and the analytical ones) have to be kept separate.
2. The citation indicator should accommodate various evaluation schemes, for example, by funding agencies. Some agencies may be interested in the top-1% (e.g., National Science Board, 2010) while others may be interested in whether papers based on research funded by a given agency perform significantly better than comparable non-funded ones (e.g., Bornmann *et al.*, 2010).
3. The indicator should allow for productivity to be taken into account. One should, for example, be able to compare two papers in the 39$^{th}$ percentile with a single paper in the 78$^{th}$ percentile (with or without weighting the differences in rank in an evaluation scheme as specified under 2.).
4. The indicator should provide the user, among other things, with a relatively straightforward criterion for the ranking (for example, a percentage of a maximum) that can then be tested for its statistical significance in relation to comparable (sets of) papers.
5. It should be possible to calculate the statistical errors of the measurement.

Using the publications of seven principal investigators at the Amsterdam Medical Center (AMC), we showed in detail how one can use percentiles and test the non-parametric differences (e.g., in SPSS) using Bonferroni corrections. In our opinion, this should have become the basis for a new "crown indicator", but we are not in the business of using indicators in evaluation practices.

The proposal by Glänzel & Schubert (2018) to return to the first-generation indicators of the mid-80s and 90s stands athwart this progression. The argument that at the aggregate level, relative indicators provide another and sometimes perhaps richer perspective does not legitimate their use in the practice of research evaluations. In a medical practice, for example, if someone deliberately used a value other than the statistically expected one for making a decision, the doctor would be held responsible for the (potentially lethal) consequences. In the rat-race for university positions and academic status, however, this collateral damage seems to be taken for granted.

In policy-making and managerial contexts, one can work with a flawed or outdated indicator so long as no alternatives are at hand (Leydesdorff, Wouters, & Bornmann, 2016). In other words, the functionality of the indicators is a pragmatic issue, and relatively independent of the validity of the results (Dahler-Larsen, 2014; cf. Hicks *et al.*, 2015). As Lehman, Jackson, & Lantrup (2006) formulated: "There have been few attempts to discover which of the popular citation measures is best and whether any are statistically reliable." Gingras (2016, at p. 76) noted that indicators without a foundation in methodology can only be explained by marketing strategies on the part of the producers.

**Perspectives for further research**

Two main problems remain when working with *MNCS* as a new crown indicator:

1. Using the mean of the (highly skewed) distribution as the expectation (Seglen, 1992). The Leiden Rankings have proceeded using percentiles (Waltman *et al*., 2012), but in many other evaluation studies *MNCS* is used based on average citation scores in Web-of-Science Subject Categories.
2. Using the Web-of-Science Subject Categories (WCs) for the delineation of the reference sets. These sets are defined at the journal level. Journals, however, are an amalgam of different subfields and therefor a poor basis for creating reference values (Opthof, 2011). WCs remain at the level of journals because the fields are defined as combinations of journals.

Pudovkin & Garfield (2002) described the method and history of how journals have been assigned Subject Categories in the *JCR*. The authors state that journals are assigned categories by "subjective, heuristic methods" (p. 1113), which the authors clarify in a footnote as follows:

> …This method is "heuristic" in that the categories have been developed by manual methods started over 40 years ago. Once the categories were established, new journals were assigned one at a time. Each decision was based upon a visual examination of all relevant citation data. As categories grew, subdivisions were established. Among other tools used to make individual journal assignments, the Hayne-Coulson algorithm is used. The algorithm has never been published. It treats any designated group of journals as one macrojournal and produces a combined printout of cited and citing journal data. (p. 1113n.)

According to the evaluation of these authors, in many fields these categories are sufficient; but they also acknowledge that "in many areas of research these 'classifications' are crude and do not permit the user to quickly learn which journals are most closely related" (p. 1113). These problems have not been diminished but have increased with the more recent expansions of the database (Leydesdorff & Bornmann, 2016).